\documentclass[a4paper,12pt,titlepage]{article}

\usepackage{amsmath,amssymb}
\usepackage{geometry}
\usepackage{hyperref}
\usepackage{braket}
\usepackage{titlesec}
\usepackage{hyperref}
\usepackage{fancyhdr}
\usepackage[dvipdfmx]{graphicx}
\usepackage{slashed}

\hypersetup{
    setpagesize=false,
    bookmarksnumbered=true,
    bookmarksopen=true,
    colorlinks=true,
    linkcolor=blue,
    citecolor=blue,
}

\begin{document}

\fancypagestyle{foot}
{
    \fancyfoot[L]{\(^{*}\)E-mail address: yakkuru$\_$111@ruri.waseda.jp}
    \fancyfoot[C]{}
    \fancyfoot[R]{}
    \renewcommand{\headrulewidth}{0pt}
    \renewcommand{\footrulewidth}{0.5pt}
}

\renewcommand{\footnoterule}{%
  \kern -3pt
  \hrule width \columnwidth
  \kern 2.6pt}


\begin{titlepage}
    \begin{flushright}
        \begin{minipage}{0.2\linewidth}
            \normalsize
            WU-HEP-20-05
        \end{minipage}
    \end{flushright}

    \begin{center}
        \vspace*{5truemm}
        \Large
        \bigskip \bigskip
        \LARGE \textbf{Local String Compactifications from Matrix Model} 
        
        \Large

        \bigskip \bigskip
        Masaki Honda$^{1,*}$ 

        \smallskip

        {\large $^{1}$ {\it Department of Physics, Waseda University, Tokyo 169-8555, Japan}} \\ 

        \bigskip \bigskip \bigskip
        
        \large \textbf{Abstract}
    \end{center}
    Matrix models are proposed as nonperturbative formulations of superstring theory. We study a concrete correspondence of the analytical result between the matrix model and the field theory. In this paper, we focus on a fuzzy sphere and a complex projective space. We show a wave functions/states correspondence of the zero modes of the Dirac operators under the 't Hooft-Polyakov monopole background on the continuous/fuzzy sphere. In addition, we propose a Laplacian for a scalar sector of the matrix model and obtain a concrete construction of it.
    \thispagestyle{foot}
    
\end{titlepage}

\baselineskip 7.5mm


\tableofcontents

\section{Introduction}
\label{1}
Superstring theory is a promising candidate for a unified theory of all forces in nature. However, superstring theory has vacua, which are degenerated infinitely. Therefore, superstring theory may have no predictions for our world, and a more fundamental theory is necessary.

Matrix models are proposed as nonperturbative formulations of superstring theory. In this paper, we consider Ishibashi-Kawai-Kitazawa-Tsuchiya (IKKT) model with some deformations~\cite{Ishibashi:1996xs,Iso:2001mg} as an extradimensional model. It is believed that the degree of freedom of spacetime and matter fields are embedded into that of matrices. Therefore, constructing a method to read out such information is important. In this paper, we mainly focus on aspects of matter fields. 

There are several attempts to analyze topological aspects like the index theorem. The index of a Dirac operator is important to confirm the realization of chiral fermions. In Ref.~\cite{Aoki:2002fq}, the authors proposed the general construction of a Ginsparg-Wilson (GW) Dirac operator on fuzzy manifolds. In addition, they proved the index theorem for a GW Dirac operator constructed by their method. In Ref.~\cite{Aoki:2003ye}, the authors showed the relationship of the Index theorem between the fuzzy sphere and the continuous sphere.

On the other hand, analytical aspects like wave functions are not obvious. In a extradimensional model, a coupling constant in four dimensions is calculated by an overlap integral of the wave functions on the extra dimensions. Therefore, it is important whether a wave function can be obtained from the result of matrix models. In addition, we have to consider scalar sectors if we keep in mind, for example, Yukawa couplings. We tend to concentrate on the discussion for fermions because of the structure of the standard model. However, we have to realize corresponding result with respect to the scalar sectors since matrix models should include corresponding structures.

In Ref.~\cite{Honda:2019bdi}, the author obtained the wave functions and the Yukawa couplings on the noncommutative magnetized torus. Identifying a matrix as an operator led us to the simple manipulation. The result was consistent with Ref.~\cite{Cremades:2004wa}. However, from the viewpoint of a regularization of field theories, we have to show such a correspondence in finite dimensional cases. 

The purpose of this paper is to show an analytical correspondence for scalar and fermion sectors between the result of the matrix models for fuzzy manifolds and that of field theoretical models by considering a concrete model.

In this paper, we will focus on a fuzzy sphere and a complex projective space discussed in Ref.~\cite{Conlon:2008qi}. According to Ref.~\cite{Conlon:2008qi}, the 't Hooft-Polyakov (TP) monopole as a magnetic flux was important for the chirality and the generation structure. In addition, the authors derived the wave functions of the fermions and the scalar fields and calculated the Yukawa coupling. On the other hand, the TP monopole on the fuzzy sphere was introduced in Ref.~\cite{Baez:1998he}. In Refs.~\cite{Aoki:2003ye,Balachandran:2005ew}, the eigenvalue problem of the GW Dirac operator were discussed based on the $su(2)$ algebra. However, there is no discussion of the correspondence between the wave functions and the states. Moreover, the eigenvalue problem of scalar sectors is not mentioned.

The organization of this paper is as follows. In Section~\ref{2}, we briefly review the IKKT model for the fuzzy sphere. In Section~\ref{3}, we review the GW algebra and the TP monopole on the fuzzy sphere. In Section~\ref{4}, we will show the correspondence between the wave functions and the states by the {\it coherent spin states}. In Section~\ref{5}, we will obtain a Laplacian, which is consistent with Ref.~\cite{Conlon:2008qi}, for the scalar sector. Section~\ref{6} contains conclusions and discussion. In the Appendix, we review the spin coherent states and discuss a simple extension of the scalar sector in the matrix model.

\section{IKKT model and fuzzy sphere}
\label{2}
\label{2.1}
In the IKKT model, the fuzzy sphere is realised as a classical solution of the variational problem. The action is defined as follows~\cite{Iso:2001mg},
\begin{align}
\label{action}
    S = \frac{1}{g^{2}} \operatorname{Tr} \left( -\frac{1}{4} [X_{i},X_{j}][X^{i},X^{j}] + \frac{2}{3} i \alpha \epsilon_{ijk} X^{i}X^{j}X^{k} + \frac{1}{2} \bar{\psi} \sigma^{i} [X_{i}, \psi] + \alpha \bar{\psi} \psi     \right),
\end{align}
where $X_{i}(i=1,2,3)$ and $\psi$ is $N \times N$ hermitian matrices, and $\psi$ is a three-dimensional Majorana spinor. $\sigma^{i}$ is Pauli matrices. The indices are contructed by the three-dimensional Euclidean metric. $\alpha$ is a dimensionfull parameter. This action is a reduced model~\cite{Eguchi:1982nm} of supersymmetric Yang-Mills with Chern-Simons and Majorana mass terms.

This action has $\mathcal{N}=1$ supersymmetry,
\begin{align*}
    \delta X_{i} = i \bar{\epsilon} \sigma_{i} \psi, \quad \delta \psi = \frac{i}{2} [X_{i},X_{j}] \sigma^{ij} \epsilon,
\end{align*}
where $\epsilon$ is a three-dimensional Majorana spinor. This symmetry implies that Chern-Simons term must have the same parameter $\alpha$ with Majorana mass term. 

Next, we consider the variational problem with respect to $X_{i}$. If we set $\psi=0$, it is 
\begin{align}
\label{eom}
    \left[X_{i},[X_{i},X_{j}] \right] = -i \alpha \epsilon_{jkl}[X_{k}, X_{l}].
\end{align}

The simplest solution is that $X_{i}$ commutes each other. However, in such a case, we can show that attractive forces act between the eigenvalues in the one-loop effective potential around this vacuum. Therefore, they do not spread, and the correspondence with the original theory does not hold. It is known that we need to consider more conditions on gauge groups~\cite{Austing:2001bd,Austing:2001pk}.

An interesting solution of eq.~(\ref{eom}) is that $X_{i}$ satisfies {\it the algebra of the fuzzy sphere}, i.e.,
\begin{align*}
    [X_{i},X_{j}] = i \alpha \epsilon_{ijk} X_{k}.
\end{align*}
If we define $X_{i} =  \alpha \hat{L}_{i}$, $\hat{L}_{i}$ satisfies the $su(2)$ algebra. Accordingly, we can consider the quadratic Casimir operator with spin $L$ ($2L+1$-dimensional) representation,
\begin{align*}
    \sum_{i=1}^{3} (X_{i})^{2} = \alpha^{2}  \sum_{i=1}^{3} (\hat{L}_{i})^{2} = \alpha^{2}L(L+1).
\end{align*}
On the matrix algebra $Mat(2L+1)$, the $su(2)$ algebra can act either on the left or on the right. In the following, we denote $\hat{L}^{L}_{i}$ and $-\hat{L}^{R}_{i}$, respectively. Both of the operators have the spectrum $L(L+1)$. In the $\alpha \rightarrow 0$, $X_{i}$ becomes usual spherical coordinate,
\begin{align*}
    X_{1} \rightarrow R \sin{\theta} \cos{\phi}, \quad X_{2} \rightarrow R \sin{\theta} \sin{\phi}, \quad X_{3} \rightarrow R \cos{\theta},
\end{align*}
where $R$ is the radius of the corresponding sphere.

Therefore, the fuzzy sphere can be realized dynamically from the action~(\ref{action}).

\section{Dirac operator and 't Hooft-Polyakov monopole on fuzzy sphere}
\label{3}

\subsection{Ginsparg-Wilson algebra and Dirac operator}
\label{3.1}

Dirac operators are fundamental objects to analyze phenomenological aspects in extradimensional models since its zero modes may correspond to the standard model particles. In continuous manifold cases, we can construct a Dirac operator from a metric or a vielbein. However, in fuzzy manifold cases, their concepts are less specific than continuous cases. 

In Ref.~\cite{Aoki:2002fq}, the authors proposed a method that is valid on any fuzzy manifolds to construct a Ginsparg-Wilson (GW) Dirac operator based on the GW algebra. In Ref.~\cite{Balachandran:2005ew}, there are discussion to complement Ref.~\cite{Aoki:2002fq}.

The GW algebra $\mathcal{A}_{\text{GW}}$ is defined as the unital $\ast$-algebra over $\mathbf{C}$ generated by $\Gamma,\Gamma'$ such as
\begin{align}
    \label{GWalg}
    \mathcal{A}_{\text{GW}} := \left<\Gamma,\Gamma^{'}: \Gamma^{2}=\Gamma^{'2}=\mathbf{1}, \Gamma^{*}=\Gamma, \Gamma^{'\ast}=\Gamma^{'}  \right>,
\end{align}
where $\ast$ is a conjugate (in the following, we interpret as the Hermitian conjugate) and an associated algebraic operation is the matrix algebra.

A GW Dirac operator $D_{\text{GW}}$ is defined as an element of $\mathcal{A}_{\text{GW}}$~\cite{Aoki:2002fq} such that 
\begin{align}
    \label{GW Dirac}
    f(a:\Gamma)D_{\text{GW}} = 1-\Gamma \Gamma^{'},
\end{align}
where $a$ is a parameter which corresponds to a lattice spacing. We assume that $f(a:\Gamma)$ has an inverse element. Since the GW Dirac operator satisfies 
\begin{align*}
    \Gamma D_{\text{GW}} + \Gamma^{'}D_{\text{GW}}=0,
\end{align*}
then we can prove the index theorem 
\begin{align*}
    \operatorname{Index}\left(D_{\text{GW}} \right) = \frac{1}{2} \operatorname{Tr} \left( \Gamma + \Gamma^{'} \right).
\end{align*}

In the following, we consider 
\begin{align*}
    D_{\text{GW}} = \frac{1}{a} \left( \Gamma - \Gamma^{'} \right), \quad S = \operatorname{Tr} \left( \bar{\Psi} D_{\text{GW}} \Psi  \right).
\end{align*}

\subsection{'t Hooft-Polyakov monopole on fuzzy sphere}
\label{3.2}
In this subsection, we consider how to introduce the 't Hooft-polyakov (TP) monopole on the fuzzy sphere. According to Ref.~\cite{Baez:1998he}, the TP monopole on the fuzzy sphere is realized as an extra angular momentum operator for the fuzzy sphere algebra. The TP monopole on the fuzzy sphere is also considered in Ref.~\cite{Aoki:2003ye}.

In the continuous manifold cases, a non-trivial vector bundle can be realized by a projection from a trivial bundle~\cite{Serre,Swan}. For the 't Hooft monopole bundle with the monopole charge $\pm N$ on a sphere, the projection operator is given by 
\begin{align*}
    &\mathcal{P}^{\pm N} = \prod_{i=1}^{N} \frac{1 \pm \vec{\sigma}^{(i)} \cdot \vec{n}  }{2 },\\
    &\vec{\sigma}^{(i)} := \underbrace{ \mathbf{1}_{2} \otimes \cdots \otimes \stackrel{i}{\vec{\sigma}} \otimes \cdots \otimes \mathbf{1}_{2} }_{N},
\end{align*}
where $n_{i}(i=1,2,3)$ is a unit vector on the sphere.

In the fuzzy sphere, the analogous operator for the monopole charge $\pm1$ is given by
\begin{align*}
    p^{(\pm 1)} =  \frac{1 \pm \vec{\sigma} \cdot \vec{x}  }{2 },
\end{align*}
where $x_{i}(i=1,2,3)$ is an element of the fuzzy sphere algebra satisfies $\sum_{i=1}^{3} (x_{i})^{2}=1$. However, this operator does not work as a projection operator since $x_{i}$ does not commute each other. On the other hand, if we consider a generator of $su(2)$ algebra $\hat{T}_{i}$ with the spectrum $T(T+1)$, then
\begin{align*}
    p^{(+1)} := \frac{1+\gamma_{\chi}}{2}, \quad \gamma_{\chi} := \frac{\vec{\sigma} \cdot \vec{\hat{T}}+1/2}{T+1/2}
\end{align*}
is an idempotent operator remarked in Refs.~\cite{CarowWatamura:1996wg,CarowWatamura:1998jn}\footnotemark[1]\footnotetext[1]{In Refs.~\cite{CarowWatamura:1996wg,CarowWatamura:1998jn}, $p^{(+1)}$ is defined as a chirality operator on the fuzzy sphere.}. If we combine $\hat{T}_{i}$ and $\frac{1}{2} \sigma_{i}$ into the generator $\hat{K}^{(+1)}_{i}:=\hat{T}_{i} + \frac{1}{2} \sigma_{i}$ with the spectrum $K(K+1),(K=T \pm \frac{1}{2})$, the operator $p^{(+1)}$ can be interpreted as the projection operator into $K=T+\frac{1}{2}$ subspace,
\begin{align*}
    p^{(+1)} = \frac{\sum_{i=1}^{3} \left( \hat{K}^{(+1)}_{i} \right)^{2} - (T-1/2)(T+1/2)  }{ (T+1/2)(T+3/2)- (T-1/2)(T+1/2)}.
\end{align*}
From this, we can construct the projection operator for the monopole charge $+N$,
\begin{align*}
    p^{(+N)}:= \frac{\prod_{K \neq K_{\text{max}}}[\sum_{i=1}^{3} \left( K^{(+N)}_{i} \right)^{2} - K(K+1)] }{\prod_{K \neq K_{\text{max}}} [K_{\text{max}}(K_{\text{max}}+1)-K(K+1)]},
\end{align*}
where $\hat{K}^{(+N)}_{i}:= \hat{T}_{i} + \sum_{i=1}^{N}\frac{1}{2} \sigma^{(i)}_{i}$ and $K_{\text{max}}=K+\frac{N}{2}$. Similarly, the projection operator for the monopole charge $-N$ can be constructed from the minimum value $K_{\text{min}}=T-\frac{N}{2}$.

Therefore, the 't Hooft-Polyakov monopole on the fuzzy sphere can be constructed by an extra angular momentum operator.

\subsection{Dirac operator}
\label{3.3}

As in the subsection~\ref{3.1}, a GW Dirac operator can be realized as an element of the GW algebra. In Ref.~\cite{Balachandran:2005ew}, the GW algebra for the TP monopole background on the fuzzy sphere is given by
\begin{align*}
    \Gamma^{\pm} := \frac{ \vec{\sigma}(\vec{\hat{L}}^{L} + \vec{\hat{T}} ) + 1/2 }{L \pm T + 1/2}, \quad \Gamma^{'}= -\frac{-\vec{\sigma} \cdot \vec{\hat{L}}^{R} + 1/2 }{L + 1/2}, \quad a= \frac{1}{\sqrt{(L+1/2)(L \pm T + 1/2)}},
\end{align*}
where $\hat{T}_{i}$ with the spectrum $T(T+1)$ corresponds to the extra angular momentum operator in the subsection~\ref{3.2}. In addition, we have to constraint ourselves into $\hat{L}^{L}_{i} + \hat{T}_{i}$ with the spectrum $(L \pm T)(L \pm T +1)$. From the definition, the square of $D_{\text{GW}}$ is given by
\begin{align*}
    (D_{\text{GW}})^{2} = \left( \vec{\hat{L}}^{L} - \vec{\hat{L}}^{R} + \vec{\hat{T}}   + \frac{1}{2} \vec{\sigma}  \right)^{2} + \frac{1}{4} - T^{2}.
\end{align*}
If we define the generator $J_{i}=\hat{L}^{L}_{i} - \hat{L}^{R}_{i} + \hat{T}_{i} + \frac{1}{2} \sigma_{i} $ with the spectrum $J(J+1)$, then $J=T-\frac{1}{2}$ is the zero mode states with the degeneracy is $2T$. This result is consistent with the index theorem for the corresponding sphere.

\section{Wave function/state correspondence}
\label{4}
As in the subsection~\ref{3.3}, the zero modes of $D_{\text{GW}}$ are written by the states of the angular momentum $J_{i}$. The topological aspect like the index is the same with the continuous sphere. However, the analytical aspect like wave function is not yet obvious. In the continuous case, wave functions are important to calculate four-dimensional coupling constants like Yukawa couplings through overlap integrals. 

In this section, we derive wave functions obtained in Ref.~\cite{Conlon:2008qi} from the zero mode states by using the coherent spin states~\cite{Radcliffe}\footnotemark[2]\footnotetext[2]{In Ref.~\cite{Iso:2001mg}, the authors considered the coherent spin states to obtain the map from matrix models to field theories with the star product.}. We review the basics of the coherent spin states in Appendix~\ref{A}.

The coherent spin state corresponding the zero mode state of $D_{\text{GW}}$ is given by
\begin{align*}
    \ket{z} := \frac{1}{N^{1/2}} \sum_{p=0}^{2T-1}  \left( \frac{(2T-1)!}{p!(2T-p-1)!}  \right)^{1/2} z^{p} \ket{p},
\end{align*}
where $N$ is a normalization factor. In our case, there are $2T$ zero mode states in the area $4 \pi R^{2}$. Therefore, we require the normalization factor satisfying
\begin{align*}
    \braket{z|z} = \frac{1}{N} (1+|z|^{2})^{2T-1} \equiv \frac{2T}{4 \pi R^{2}},
\end{align*}
and hence the normalized state is written as
\begin{align*}
    \ket{z} := \frac{1}{(1+|z|^{2})^{\frac{2T-1}{2}}} \sum_{p=0}^{2T-1}  \left(\frac{1}{4 \pi R^{2}} \cdot \frac{(2T)!}{p!(2T-p-1)!}  \right)^{1/2} z^{p} \ket{p}.
\end{align*}
Then, we can derive the wave functions obtained in Ref.~\cite{Conlon:2008qi} by the inner product
\begin{align*}
    \braket{p|z} =  \left(\frac{1}{4 \pi R^{2}} \cdot \frac{(2T)!}{p!(2T-p-1)!}  \right)^{1/2} \frac{z^{p}}{(1+|z|^{2})^{\frac{2T-1}{2}}} \equiv \Psi^{p}_{2T}.
\end{align*}

On the other hand, we have to define the integral measure $m(|z|^{2}) \geq 0$ to satisfy
\begin{align}
\label{measure1}
    \int d^{2}z~m(|z|^{2}) \ket{z} \bra{z} = \sum_{p=0}^{2T-1} \ket{p} \bra{p} = \mathbf{1}.
\end{align}
Eq.~(\ref{measure1}) requires $m(|z|^{2})=4R^{2} \frac{1}{(1+|z|^{2})^{2}}$, and the integral
\begin{align*}
    4R^{2} \int d^{2}z \frac{1}{(1+|z|^{2})^{2}}
\end{align*}
is just the integral over the sphere with the metric
\begin{align*}
    ds^{2} = \frac{4R^{2}}{(1+|z|^{2})^{2}} dz d\bar{z}.\\
\end{align*}

\section{Scalar sector with 't Hooft-Polyakov monopole on fuzzy sphere}
\label{5}
So far, we considered the fermionic sector on the fuzzy sphere with the TP monopole. On the other hand, We have to mention about a scalar sector. In Ref.~\cite{Conlon:2008qi}, there is no scalar zero modes while the magnetic flux is zero. For the lowest modes, the wave function $\phi(z,\bar{z})$ and its mass $m^{2}$ are given as
\begin{align*}
    \phi(z,\bar{z}) =  \left(\frac{1}{4 \pi R^{2}} \cdot \frac{(2T+1)!}{p!(2T-p)!}  \right)^{1/2} \frac{z^{p}}{(1+|z|^{2})^{\frac{2T}{2}}} \equiv \Psi^{p}_{2T+1}, \quad m^{2}=\frac{2T}{2R^{2}},
\end{align*}
in our notation. The difference from fermions is the magnetic flux $2T \to 2T+1$ $(T -\frac{1}{2} \to T)$. From this, we expect the Laplacian is given by
\begin{align*}
    R^{2} \Delta_{\text{GW}} &= \vec{\mathcal{K}}^{2} - T(T+1) + T \\
    &=\vec{\mathcal{K}}^{2} -T^{2},
\end{align*}
where $\mathcal{K}_{i}= \hat{L}^{L}_{i} - \hat{L}^{R}_{i}+ T_{i}$ with the spectrum $K(K+1)$ and the corresponding states are labeled by $K=T$. However, according to Appendix~\ref{B}, the simple extension is not appropriate since there is no $-T^{2}$ term.

To construct the expected Laplacian, we consider
\begin{align*}
    \Gamma^{\pm}_{i} := \frac{ \hat{L}^{L}_{i} + \hat{T}_{i}   }{L \pm T }, \quad \Gamma^{'}_{i}= -\frac{ -\hat{L}^{R}_{i} }{L}, \quad a=\frac{1}{ \sqrt{L(L \pm T)} },
\end{align*}
where we also constraint ourselves into $\hat{L}^{L}_{i} + \hat{T}_{i}$ with the spectrum $(L \pm T)(L \pm T +1)$. These are $\Gamma^{\pm}$ and $\Gamma^{'}$ without the spinor and the curvature contributions ($\sigma_{i}$ and $1/2$ factor, respectively). Then, $R^{2}\Delta_{\text{GW}}$ can be realized by
\begin{align*}
    R^{2}\Delta_{\text{GW}} = (D_{\text{GW},i})^{2}, \quad D_{\text{GW},i}:= \frac{1}{a} \left(  \Gamma^{\pm}_{i} - \Gamma^{'}_{i}\right).
\end{align*}
This is a natural construction since the scalar sector does not have the spinor index and the curvature does not affect the scalar sector.

\section{Conclusions}
\label{6}
In this paper, we have studied the reconstruction of the local model proposed in Ref.~\cite{Conlon:2008qi} based on the matrix model. In section~\ref{4}, we reconstructed the zero mode wave functions which are derived in Ref.~\cite{Conlon:2008qi} from zero mode states under the TP monopole background on the fuzzy sphere by using the coherent spin states. In section~\ref{5}, we proposed the Laplacian, which has the appropriate eigenvalue, and obtained the concrete construction.

For the scalar sector, our construction is interesting. Our construction may be valid for other fuzzy manifolds since the method of the GW algebra is valid on any fuzzy manifolds. However, in general, identifying the spinor and curvature contributions are difficult. Therefore, it is important to construct other models where we can identify them.

We are also interested in the intersecting model of the fuzzy spheres. In Refs~\cite{Nishimura:2013moa,Aoki:2014cya}, the authors realized chiral fermions by using the numerical calculation. On the other hand, for the fuzzy sphere, the analysis from the viewpoint of angular momentum was effective. Therefore, we think that there is scope for analytical discussion like Ref~\cite{Chatzistavrakidis:2011gs}, but it is our future work.

\section*{Acknowledgments}
The author would like to thank H. Abe and G. Ishiki for helpful comments.

\def\thesubsection{\Alph{subsection}}
\setcounter{subsection}{0}

\renewcommand{\theequation}{A.\arabic{equation}}
\setcounter{equation}{0}

\section{Appendix}
\label{app}
\subsection{Coherent spin states}
\label{A}

In Ref.~\cite{Radcliffe}, coherent spin states are defined as an analogy of coherent states of the harmonic oscillator. We start from the angular momentum operator $\hat{S}_{i}$ with the spectrum $S(S+1)$. The ground state is the highest spin state,
\begin{align*}
    \hat{S}_{z} \ket{S:S} = S \ket{S:S}.
\end{align*}

In addition, the state $\ket{p}$ is defined by
\begin{align*}
    (\hat{S}_{-})^{p} \ket{S:S} = \left( \frac{p! (2S)!}{(2S-p)!} \right)^{1/2} \ket{p} \quad (0 \leq p \leq 2S),
\end{align*}
where $\hat{S}_{-}:=\hat{S}_{x}-i\hat{S}_{y}$. The coherent spin states are defined by 
\begin{align*}
    \ket{z} := \frac{1}{N^{1/2}} \exp{ \left( z \hat{S}_{-} \right) } \ket{S:S} = \frac{1}{N^{1/2}} \sum_{p=0}^{2S} \left( \frac{(2S)!}{p!(2S-p)!}  \right)^{1/2} z^{p} \ket{p},
\end{align*}
where $N$ is a normalization factor\footnotemark[3]
\footnotetext[3]{In section~\ref{4}, we substitute the normalization factor and the integral measure because of the degeneracy of the zero mode states.}. Since the inner product is
\begin{align*}
    \braket{z|z}=\frac{1}{N} \sum_{p=1}^{2S} \frac{(2S)!}{p!(2S-p)!} |z|^{2p} = \frac{1}{N} (1 + |z|^{2})^{2S} \equiv 1,
\end{align*}
then $N$ should be $(1+|z|^{2})^{S}$.

Moreover, we have to define the integral measure\footnotemark[3] $m(|z|^{2}) \geq 0$ to satisfy
\begin{align}
\label{hu}
\int d^{2}z\ m(|z|^{2}) \ket{z} \bra{z} = \sum_{p=1}^{2S} \ket{p} \bra{p} = \mathbf{1}.
\end{align}
Eq.~(\ref{hu}) requires $m(|z|^{2})= \frac{2S+1}{\pi} \frac{1}{(1+|z|^{2})^{2}}$.

\subsection{Action for scalar sector}
\label{B}
In this subsection, we consider a simple extension of the action for a scalar sector. Usually, an action of the matrix model for a scalar sector without the TP monopole on the fuzzy sphere is written by
\begin{align*}
    S=\operatorname{Tr}\left( [\hat{L}_{i}, \Phi] [\hat{L}_{i}, \Phi] \right) = \operatorname{Tr}\left( -\Phi [\hat{L}_{i},  [\hat{L}_{i}, \Phi]] \right) = \operatorname{Tr}\left( -\Phi (\hat{\mathcal{L}}_{i})^{2} \Phi \right),
\end{align*}
where $\Phi$ is a Hermitian matrix and $\hat{\mathcal{L}}_{i} = [\hat{L}_{i}, \cdot]=\hat{L}^{L}_{i} - \hat{L}^{R}_{i}$ is the $su(2)$ generator with the spectrum $\mathcal{L}(\mathcal{L}+1),\mathcal{L} \in \{0, \dots, 2L\}.
$

The TP monopole on the fuzzy sphere can be introduced by substituting $\hat{\mathcal{K}}_{i} := \hat{\mathcal{L}}_{i} + \hat{T}_{i}$ for $\hat{\mathcal{L}}_{i}$. Therefore, the action of the matrix model for a scalar sector under the TP monopole background on the fuzzy sphere is written by
\begin{align*}
    S=\operatorname{Tr}\left( -\Phi (\hat{\mathcal{K}}_{i})^{2} \Phi \right) = \operatorname{Tr}\left( -\mathcal{K}(\mathcal{K}+1) \Phi \Phi \right).
\end{align*}
As we mentioned in the subsection~\ref{3.3}, we need to consider the projection of the spectrum of $\hat{L}^{L}_{i}+\hat{T}_{i}$ into $(L \pm T)(L \pm T+1)$. This implies that $\mathcal{K}$ runs from $T$ to $2L \pm T$.

\bibliographystyle{prsty}

\begin{thebibliography}{99}

\bibitem{Ishibashi:1996xs} 
  N.~Ishibashi, H.~Kawai, Y.~Kitazawa and A.~Tsuchiya,
  Nucl.\ Phys.\ B {\bf 498}, 467 (1997)
  [hep-th/9612115].
  
\bibitem{Iso:2001mg} 
  S.~Iso, Y.~Kimura, K.~Tanaka and K.~Wakatsuki,
  Nucl.\ Phys.\ B {\bf 604}, 121 (2001)
  [hep-th/0101102].
  
\bibitem{Aoki:2002fq} 
  H.~Aoki, S.~Iso and K.~Nagao,
  Phys.\ Rev.\ D {\bf 67}, 085005 (2003)
  [hep-th/0209223].
  
\bibitem{Aoki:2003ye} 
  H.~Aoki, S.~Iso and K.~Nagao,
  Nucl.\ Phys.\ B {\bf 684}, 162 (2004)
  [hep-th/0312199].
  
\bibitem{Honda:2019bdi} 
  M.~Honda,
  JHEP {\bf 1904}, 079 (2019)
  [hep-th/1901.00095].
  
\bibitem{Cremades:2004wa} 
  D.~Cremades, L.~E.~Ibanez and F.~Marchesano,
  JHEP {\bf 0405}, 079 (2004)
  [hep-th/0404229].
  
\bibitem{Conlon:2008qi} 
  J.~P.~Conlon, A.~Maharana and F.~Quevedo,
  JHEP {\bf 0809}, 104 (2008)
  [hep-th/0807.0789].
  
\bibitem{Baez:1998he} 
  S.~Baez, A.~P.~Balachandran, B.~Ydri and S.~Vaidya,
  Commun.\ Math.\ Phys.\  {\bf 208}, 787 (2000)
  [hep-th/9811169].
  
\bibitem{Balachandran:2005ew} 
  A.~P.~Balachandran, S.~Kurkcuoglu and S.~Vaidya,
  Singapore, Singapore: World Scientific (2007) 191 p.
  [hep-th/0511114].
  
\bibitem{Eguchi:1982nm} 
  T.~Eguchi and H.~Kawai,
  Phys.\ Rev.\ Lett.\  {\bf 48}, 1063 (1982).
  
\bibitem{Austing:2001bd} 
P.~Austing and J.~F.~Wheater,
``The Convergence of Yang-Mills integrals,''
JHEP {\bf 0102} (2001) 028.
[hep-th/0101071].

\bibitem{Austing:2001pk} 
P.~Austing and J.~F.~Wheater,
``Convergent Yang-Mills matrix theories,''
JHEP {\bf 0104} (2001) 019.
[hep-th/0103159].
  
  \bibitem{Serre}
  J-P.~Serre,
  Annals.\ Math.,\ {\bf 61}, 197-278 (1955).
  
  \bibitem{Swan}
  R.~G.~Swan,
  Trans.\ Am.\ Math.\ Soc.,\ {\bf 10}, 264-277 (1962).
  
\bibitem{CarowWatamura:1996wg} 
  U.~Carow-Watamura and S.~Watamura,
  Commun.\ Math.\ Phys.\  {\bf 183}, 365 (1997)
  [hep-th/9605003].
  
\bibitem{CarowWatamura:1998jn} 
  U.~Carow-Watamura and S.~Watamura,
  Commun.\ Math.\ Phys.\  {\bf 212}, 395 (2000)
  [hep-th/9801195].
  
  \bibitem{Radcliffe}
  J.~M.~Radcliffe, 
  J.\ Phys.\ A:\ Gen.\ Phys. {\bf 4}, 313 (1971).
  
\bibitem{Nishimura:2013moa} 
  J.~Nishimura and A.~Tsuchiya,
  JHEP {\bf 1312}, 002 (2013)
  [hep-th/1305.5547].
  
\bibitem{Aoki:2014cya} 
  H.~Aoki, J.~Nishimura and A.~Tsuchiya,
  JHEP {\bf 1405}, 131 (2014)
  [hep-th/1401.7848].
  
\bibitem{Chatzistavrakidis:2011gs} 
  A.~Chatzistavrakidis, H.~Steinacker and G.~Zoupanos,
  JHEP {\bf 1109}, 115 (2011)
  [hep-th/1107.0265].
  
 

\end{thebibliography}

\end{document}